# When Liquid Meets Frequency-Selective Rasorber: Wideband and Switchable 3-D Frequency-Selective Rasorber

Xiangkun Kong, *Member, IEEE*, Xuemeng Wang, Xin Jin, Weihao Lin, Lingqi Kong , Shunliu Jiang and Lei Xing, *Member, IEEE*

*Abstract*—In this paper, a switchable 3-D frequency selective rasorber (FSR) with wide absorption bands without lumped components or commercial magnetic absorbers is presented and investigated. The absorption path is constructed by embedding a hybrid liquid microwave absorber (MA) inside a parallel plate waveguide (PPW) to create an extra-wide absorption band. A reflection layer based on water is placed behind the FSR to realize the reconstruction from FSR to a band-notched absorber (BNA) by controlling the presence or absence of water. The liquid-based absorber is firstly analyzed by a multimode dielectric resonant circuit and the fundamental operating principle of the FSR is demonstrated with the help of an equivalent circuit model (ECM). A design example is provided, fabricated, and measured and it exhibits a passband at 5.10 GHz with a transmission bandwidth of 18.5% for less than 3 dB insertion loss and fractional bandwidth of 146.8% with reflectivity less than -10 dB in FSR mode. In BNA mode, it has a minimum return loss of 0.72 dB and a good absorption band from 2.5 to 4.6 GHz and 5.7 to 16.5 GHz. Good agreements among circuit analysis, simulation results, and measurement results are finally obtained. The switchable rasorber can be applied in a shared-aperture antennas system to convert a broadband stealth radome into a BNA.

## I. INTRODUCTION

Frequency selective rasorber (FSR), is a kind of structure with a low insertion transmission window in a broad absorption band[1]-[3]. In some reported studies, FSR is also termed as "absorptive frequency selective transmission [4]-[9]", "absorptive frequency selective surface [10]-[13]" and "frequency selective radome [15]". Due to its peculiar absorptive and filtering quality, it is of great significance in many applications, such as reducing RCS of antenna systems, improving communication security and reducing crosstalk between subsystems, and has attracted more and more attention in recent years.

Numerous FSR designs have been reported in the literature, during the past decade [2]-[28]. According to the published literature, the design idea of FSR can be summarized into the following two kinds: 1) cascading 2-D FSS layers [2]-[3], [9]-[17] and 2) utilizing the 3-D FSS design concept [4]-[8], [18]-[28]. The former takes its inspiration from the Salisbury screens. The specific method is to cascade a lossy layer with a lossless bandpass FSS at a quarter wavelength distance corresponding to the transmission frequency point [3]. In terms of 3-D FSR, the general strategy is to use a transmission line to create an independent transmission and absorption channel [24]. On the implementation method, the transmission line structure could be parallel plate waveguide (PPW) [24], slot line or microstrip line [28].

Wider absorption bandwidth and versatile features is one of the hotspots of FSR research. To the best of our understanding, the 2-D FSR mainly absorbs the incident electromagnetic waves in three ways: 1) lumped resistance load in lossy layer [3], 2) high-impedance surface [15] and 3) water [29]. From the standpoint of equivalent circuit model (ECM), the first two methods commonly introduce resistance capacitance inductance (RLC) series resonant circuits. Nonetheless, the number of series resonators that can be integrated on a lossy layer is restricted, resulting in a limited absorption bandwidth when the thickness of FSR stays the same.

Early 3-D FSR can be regarded as a hybrid of 3-D FSS and absorber, which builds the spectrum of absorption-transmission-absorption (ATA) through lumped resistances and cavity modes [18]-[27]. In [26], a 3-D FSR, based on cavity modes and lumped resistance, was proposed for single and dual polarization applications using PPW structure. However, this method suffers from limited absorption bandwidth. [23] designed and fabricated a 3-D type utilizing thin wideband magnetic material that achieved a fractional bandwidth (FBW) of 129.8%. Meanwhile, the thickness of the FSR was calculated to be 0.106 $\lambda_L$, where the $\lambda_L$ is the free-space wavelength at the lowest frequency of absorption band. Furthermore, the technique of inserting a series lumped L-C circuit or a bandstop FSS in front of the absorption channel that "turns off" the absorption channel at the transmission frequencies. This approach decouples the transmission and absorption channels at the transmission frequencies, resulting

This work was supported in part by National Natural Science Foundation of China under Grant 62071227, in part by Natural Science Foundation of Jiangsu Province of China under Grant BK20201289, in part by Open Research Program in China's State Key Laboratory of Millimeter Waves under Grant K202027 in part by the Postgraduate Research & Practice Innovation Program of Jiangsu Province under Grant SJCX20_0070 and in part by the Fundamental Research Funds for the Central Universities under Grant kfjj20200403.

The authors are with the College of Electronic and Information Engineering, Nanjing University of Aeronautics and Astronautics, Nanjing, 211106, China (e-mail: xkkong@nuaa.edu.cn).

in a transmission window with ultra-low insertion loss. Meanwhile, instead of using a magnetic substance, a plausible technique for achieving a greater absorption bandwidth is to load a wider MA.

Water, as one of the most generous and accessible materials on earth, is widely used in the design of microwave devices [30]-[34]. In the microwave region, water's dielectric permittivity exhibits significant dispersion property, which is well characteristic of the Debye formula [34]. Due to the relatively high real part of permittivity and mobility, water is often designed as a dielectric resonator antenna and reconfigurable antenna. Furthermore, due to the high dielectric loss, water has several applications in the domain of microwave absorber, particularly for wideband and optically transparent MA. Compared to circuit analog absorber and high impedance surface-based absorber, a deliberately designed water-based MA can excite multiple modes of dielectric resonance, resulting in a wider absorption bandwidth. The characteristic effective in forming ultrawideband (UWB) microwave absorption is what we expected in the design of FSR. However, due to the information available to the author, all water-based MA structures typically have a metal or ITO backing plate at the bottom. In [29], a water-based FSR with a transmission band above the absorption band was proposed by etching gaps in the underlying metal. Nonetheless, it suffers from a single absorption band and narrow transmission bandwidth.

In this paper, a switchable 3-D FSR with ultra-wide absorption band without lumped components or commercial magnetic absorbers is presented and investigated. To create an extra-wide absorption band, the absorption path is developed by embedding a hybrid liquid MA inside a PPW. By manipulating the presence or absence of water, a water-based reflection layer is placed below the FSR to realize the reconstruction from FSR to band-notched absorber (BNA). A multimode dielectric resonant circuit is used to analyze the liquid-based absorber first, and an ECM is used to demonstrate the FSR's fundamental operating principle. A design example is provided, fabricated, and measured and it exhibits a passband at 5.07 GHz with a transmission bandwidth of 18.5% for less than 3 dB insertion loss and fractional bandwidth of 146.8% with reflectivity less than -10 dB in ATA mode. In BNA mode, it has a minimum return loss of 0.7 dB and a good absorption band from 2.5 to 4.6GHz and 5.7 to 16.5GHz. Finally, good agreement among circuit analysis, simulation findings, and measurement results is achieved.

The structure of this article is organized as follows. Section II describes the structure of our 3-D FSR. Section III explains the operating principle in detail based on an equivalent circuit model. The fabrication and measurement of our proposed FSR are introduced in Section IV, while the performance and comparison are also discussed. Finally, the concluding remarks are given in Section V.

## II. DESCRIPTION OF THE FSR

Fig. 1 illustrates the configuration of our proposed FSR, which contains 3 × 3 unit cells for a perspective view of conceptual illustration. The polarization direction of the incoming plane wave is vertical for this design as shown in Fig. 1. The construction details and side view of the unit cell are displayed in Fig. 2. The unit cell periods along the $x$ and $y$ directions are denoted to P and H, respectively. Each unit cell consists of two PPW structures to provide the absorption and transmission channels independently. To generate an extra-wide absorption band, a Propylene glycol (PG) aqueous solution MA is embedded into a PPW structure. And a mender line FSS that the thickness of its substrate is 0.5 mm in shunt connection with the absorber is used to replace lumped components and to achieve a low insertion loss at the passband. The hybrid liquid MA is composed of a resin 3D printing material ($\varepsilon_r^*=2.8\text{-}j0.0318$), a metal backplate, and PG aqueous solution measured by DAK dielectric parameter measuring instrument. To improve impedance matching, an air gap is inserted into absorption channel for a wider absorption bandwidth which has been demonstrated in [23]. The transmission path is constructed from a substrate path based on a step impedance resonator (SIR) with a metallic via in the middle and a 3-D printed container, which holds water, acts as a reflective layer. To reduce the length of the unit cell in the wave vector direction, SIR and dielectric substrate are adopted; Meanwhile, it's worth noting that some air columns were implanted in the dielectric substrate to minimize its equivalent dielectric permittivity.

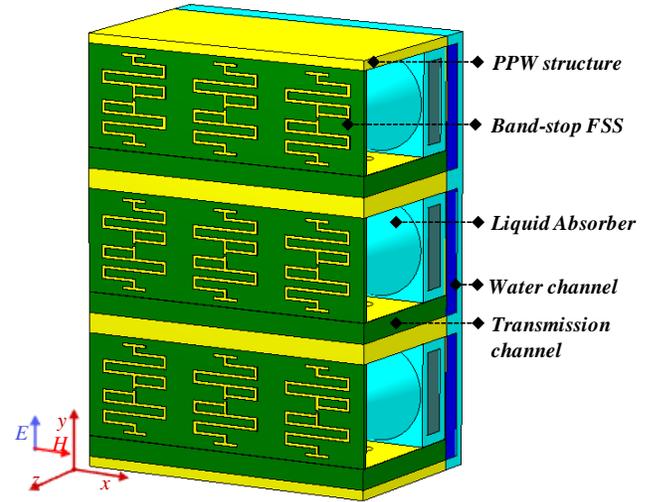

Fig. 1. Perspective view of the proposed FSR (3 × 3 unit cells for concept illustration).

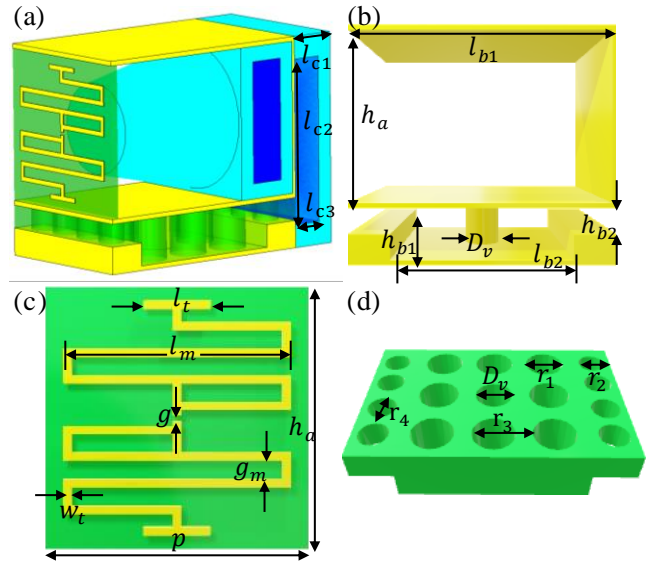

Fig. 2. Physical dimensions of a unit cell. (a) 3-D view of the FSR. (b) PPW structure. (c) Meander line band-stop FSS. (d) Dielectric substrate inserted in transmission channel. (Physical dimensions: $p = 10$, $h_a = 10$, $l_{b1} = 15.7$, $l_{b2} = 10$, $h_{b1} = 3$, $h_{b2} = 1.5$, $Dv = 1.7$, $l_t = 2.5$, $l_m = 8.5$, $g = 0.2$, $w_t = 0.3$, $g_m = 0.7$, $r_1 = 2.4$, $r_2 = 1.7$, $r_3 = 3.3$, $r_4 = 2.2$, $l_{c1} = 3$, $l_{c2} = 11$, $l_{c3} = 2$. All length units are in millimeters.

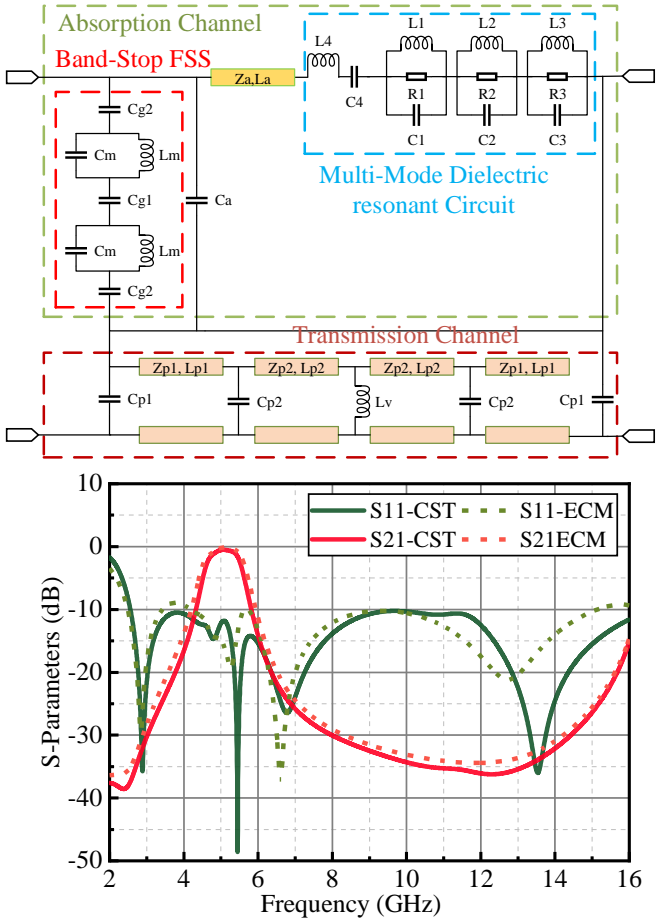

Fig. 3. Equivalent circuit model of the proposed FSR in ATA mode associated its simulated scattering parameters. ($C_g = 0.127$ pF, $C_{g1} = 0.0950$ pF, $L_m = 6.40$ nH, $C_m = 0.0785$ pF, $C_a = 2$ fF, $Z_a = 377$ Ω, $L_a = 39°@10$ GHz, $R_1 = 393.5$ Ω, $R_2 = 395.1$ Ω, $R_3 = 291.6$ Ω, $C_1 = 0.383$ pF, $C_4 = 0.0719$ pF, $C_3 = 0.0326$ pF, $C_4 = 1.68$ μF, $L_1 = 8.07$ nH, $L_2 = 6.35$ nH, $L_3 = 1.65$ nH, $C_{p1} = 0.017$ pF, $C_{p2} = 0.006$ pF, $L_v = 0.346$ nH, $Z_{p1} = 41.8$ Ω, $Z_{p2} = 81.0$ Ω, $L_{p1} = 31°@7.5$ GHz, $L_{p2} = 67°@7.5$ GHz.

## III. OPERATING PRINCIPLE

### A. Equivalent circuit model

The ECM of the FSR associated with simulated scattering parameters concerning its full-wave simulation results was established and shown in Fig. 3 to illustrate the operation mechanism. It consists of two mutually independent electromagnetic wave propagation paths known as the absorption and transmission channels, respectively. The corresponding part of each PPW in the structure is indicated by an equivalent transmission line (TL). The ECM in the absorption channel is made up of three parts: Band-Stop FSS, a Multi-Mode Dielectric Resonant Circuit, and a transmission line with $Z_a$ and $L_a$ representing characteristic impedance and electric length, respectively. The multi-mode dielectric resonant circuit is used to describe the liquid MA showed in Fig. 1 that it was inspired by the ECM of a dielectric resonant antenna (DRA). The next section will explain how to set up and examine the ECM of the liquid MA. The transmission channel consists of a step impedance resonator with a center loaded inductor $L_v$. The inductor $L_v$ is the inductance of the metalized via hole, which plays a role in generating second-order bandpass response. Furthermore, $C_a$ and $C_{p1}$ denote the distributed capacitance between PPW and free space of absorption channel and transmission channel, respectively. In the transmission channel, $C_{p2}$ represents the discontinuity capacitance at the interface between the high and low impedance regions. Moreover, how to estimate the $C_a$, $C_{pl}$, $L_v$, and $C_{p2}$ have been reported in [24]. $L_m$, $C_m$, $C_{g1}$ and $C_{g2}$ are obtained by curve-fitting the simulated impedance results of the band-stop FSS.

### B. Hybrid liquid microwave absorber

Water, due to its dispersive permittivity and strong dielectric loss, has been extensively studied and exploited as a wideband, flexible and transparent MA. However, because of the impedance mismatch with free space induced by its larger permittivity, it is difficult to realize the absorption of lower frequency electromagnetic wave using a water-based microwave absorber. PG was used in this study to reduce the permittivity of pure water while maintaining its dielectric loss property. The schematic view of the hybrid liquid MA used in this article is shown in Fig. 4 (a), while the complex relative permittivity ($\varepsilon_r^* = \varepsilon_r' - j\varepsilon_r''$) of the PG aqueous solution (50% concentration) is plotted in Fig. 4 (b). As shown in Fig. 4 (a), the 50% PG aqueous solution has lower real permittivity in the frequency range from 2 to 14 GHz and higher imaginary permittivity in 1-6 GHz compared to the pure water under the temperature of 20℃. According to impedance matching theory, this is beneficial to expand the low-frequency absorbing capacity of water-based MAs.

Fig. 4 shows the schematic of the hybrid liquid MA. The MA is made up of a cone-shaped liquid block and its container, which is built up resin 3D-printing material, as shown in Fig. 2(b). Meanwhile, the container is backed by a copper ground plane. The frequency dispersion of the complex permittivity is primarily responsible for the broadband of absorption, which may be understood using an equivalent circuit model. However, no ECM has been reported in the published literature to our best knowledge.

In this section, a multi-mode dielectric resonator circuit was used to analyze the operation mechanism in physical insight. This method is inspired by the well-known concept of dielectric resonator antenna (DRA) [35]. To begin with, both water-based absorber and DRA rely on dielectric resonance to achieve their respective goals (energy radiation or electromagnetic wave absorption). The difference is that DRA gets its energy from the feed, whereas the MA obtains its energy from free space. According to the reciprocal principle, the equivalent circuit model of DRA might be used to characterize the operation mechanism of water-based MA.

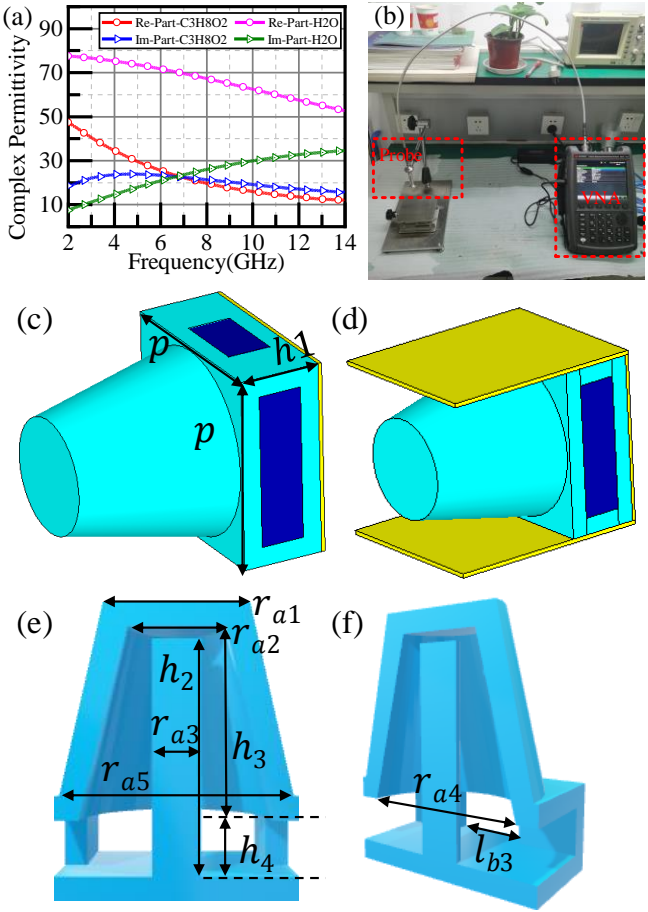

Fig. 4. Unit cell of liquid microwave absorber. (a) Complex dielectric parameters of pure water (Debye model at 20°C) and PG at 20°C. (b) The liquid dielectric parameters test system and test environment. (c) 3-D view of the liquid microwave absorber. (d) Liquid microwave absorber in PPW structure. (e) and (f) Section view of the unit cell of liquid microwave absorber. (Physical dimensions: $p = 10$, $r_{a1} = 3$, $r_{a2} = 4$, $r_{a3} = 2$, $r_{a4} = 8$, $r_{a5} = 9.5$, $h_1 = 4.3$, $h_2 = 9.8$, $h_3 = 8$, $h_4 = 2.3$. All length units are in millimeters.)

According to the foster's reactance theorem, any lossless 1-port's reactance strictly increases with frequency [36]. The input impedance of such a passive 1-port can be represented in the so-called first Foster form by a series combination of one capacitor, one inductor and a given number of parallel L-C elements. In this description, the series capacitor is used to represent an infinite impedance for $lim \, \omega \to 0$, the series inductor to represent an infinite impedance for $lim \, \omega \to \infty$, and the L-C elements are used to represent poles resonances of the 1-port. Since the provided absorber is a lossy structure, the initial Foster form must be modified to properly describe the absorber. To account for the losses, the parallel L-C elements in the first Foster form are connected with a parallel resistor, resulting in R-L-C elements. The necessary number of R-L-C elements can be determined by observing the modes that are effectively generated. Fig. 5 illustrates the ECM of the MA proposed in this article. It consists of three groups of parallel L-C-R elements ($R_i, L_i, C_i$, and $i = 1,2,3$) and a series L-C element ($L_0, C_0$). To determine the parameters of 11 unknown values faster, we used the parameters: ($Q_i, k_i, f_i, X_{L0}, X_{C0}$, and $i = 1,2,3$) to represent the value of resistance, inductance and capacitance. The $Q_i, k_i, f_i$ represent unload $Q$ factor, the coupling coefficient of the $ith$ mode and the resonant frequency of the $ith$ parallel R-L-C elements respectively. The $X_L, X_C$ are the reactance of $L_0, C_0$ respectively. The input impedance of ECM can be calculated by

$$Z_{in}(f) = j\omega L_0 - j(\omega C_0)^{-1} + \sum_{i=1}^{3} \frac{1}{R_i^{-1} + j\omega L_i - j(\omega C_i)^{-1}}, i = 1,2,3,\ldots \quad (1)$$

where $\omega = 2\pi f$ is operating angular frequency. Furthermore, without considering cross-polarization reflection, the reflection coefficient of the input port can be obtained by

$$\Gamma = \frac{Z_{in}(f) - Z_0}{Z_{in}(f) + Z_0}, \quad (2)$$

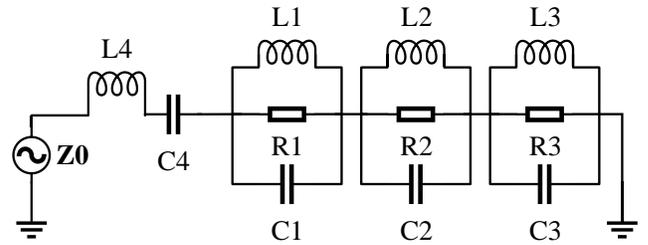

Fig. 5. Multimode dielectric resonator equivalent model of the liquid absorber based on generalized foster law ($Z0 = 377 \, \Omega$, other parameters are already in the abstract of Fig. 3.

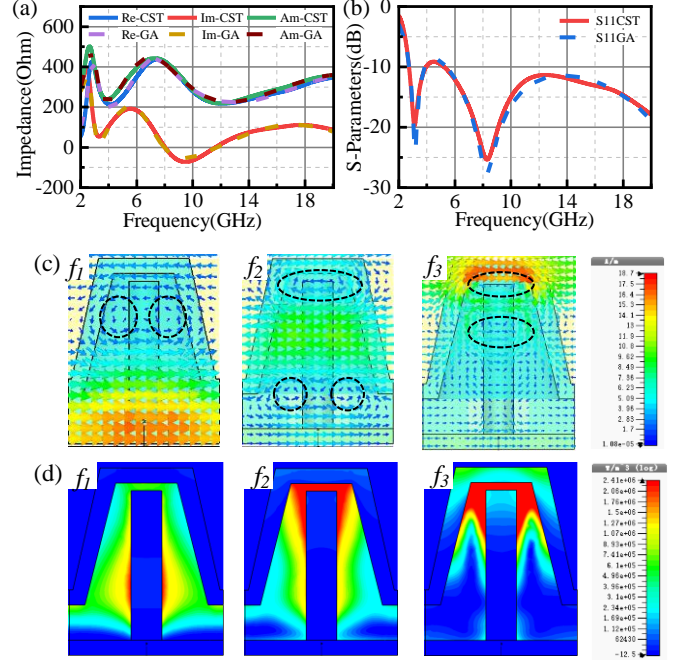

Fig. 6. H-field energy density, comparison of input impedance, scattering parameters simulation results (a) comparison of input impedance (real part, imaginary part and amplitude) of EM simulation (CST) with ECM optimized by genetic algorithm (GA) (b) power loss density distribution. ($f_1 = 2.86 \, GHz, f_2 = 7.45 \, GHz, f_3 = 21.7 \, GHz$).

Where $Z_0 = 377 \, \Omega$ is the characteristic impedance in free space. To quickly obtain the parameter search range of the optimization algorithm, $Q_i, k_i, f_i, X_L, X_C$ were used to calculate the input impedance and reflection coefficient. The impedance of each parallel R-L-C resonator could be described by

$$Z_i(f) = \frac{k_i R_0}{1 + jQ_i \xi_i(f)}, \qquad i = 1,2,3 \qquad (3)$$

where $\xi_i(f) = \frac{f}{f_i} - \frac{f_i}{f}$. The reactance of $L_0, C_0$ are described by $X_L(f) = X_{L0} \frac{f}{f_0}$ and $X_C(f) = X_{C0} \frac{f}{f_0}$, respectively. Here, $f_0$ is the center operation frequency. So, the real part, imaginary part and the amplitude of the ECM's input impedance can be expressed as

$$Z_{in}(f) = jX_L(f) + jX_C(f) + \sum_{i=1}^{3} Z_i, \qquad i = 1,2,3 \quad (4)$$

$$\begin{cases} R_{in}(f) = \sum_{i=1}^{3} \frac{k_i R_0}{1 + (Q_i \xi_i(f))^2} \\ X_{in}(f) = X_L(f) + X_C(f) - \sum_{i=1}^{3} \frac{k_i R_0 Q_i \xi_i(f)}{1 + (Q_i \xi_i(f))^2} \end{cases}, \quad i = 1,2,3 (5)$$

respectively. To obtain the 11 unknow parameters: $Q_i, k_i, f_i, X_{L0}, X_{C0}$, and $i = 1,2,3$. An objective function $G$ is defined as the square of the difference between the ECM and full-wave simulation data of input impedance as given in (6). By minimizing $G$, a set of optimized values for the 11 unknown parameters are obtained. Furthermore, to evaluate the accuracy of the proposed ECM, an averaged error rate (AER) is defined in (7): the smaller the AER, the better the accuracy.

$$G = \sum_{i=1}^{N} [Re\,(Z_i^{simu} - Z_i^{fit})]^2 + [Im\,(Z_i^{simu} - Z_i^{fit})]^2 \quad (6)$$

$$AER = \frac{1}{N} \sum_{i=1}^{N} \left| \frac{Z_i^{simu} - Z_i^{fit}}{Z_i^{simu}} \right| \times 100\% \qquad (7)$$

Where $i$ is the number of the full-wave simulation input impedance of the proposed hybrid-liquid MA. *Re* and *Im* are the real and imaginary parts of input impedance, respectively. $Z_i^{simu}$ and $Z_i^{fit}$ represent the full-wave simulation of liquid MA and *ith* ECM result of input impedance.

Comparisons have been done to assess the ECM's accuracy across the frequency band. In both full-simulation and ECM, the real parts, imaginary parts, and amplitude of the input impedance are shown in Fig. 6 (a). The S11 is plotted in Fig. 6 (b). It is observed the predicted ECM displays excellent agreement with the full-wave simulation results in the frequency range of 2 to 20GHz. Meanwhile, the AER is 2.71%, calculated throughout the frequency range 2-20GHz, indicating

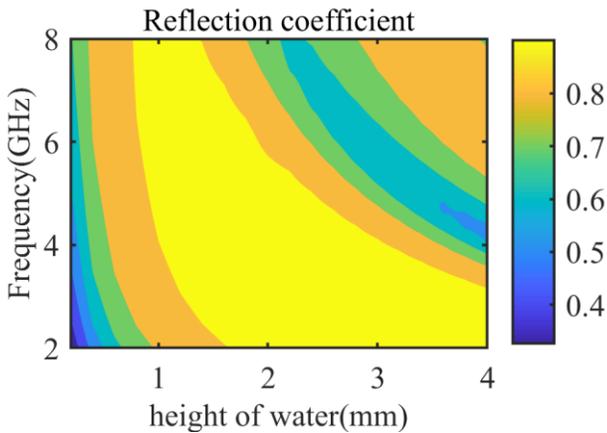

Fig. 7. (a) Fabricated prototype of the proposed FSR and (b) its simulated and measured S-parameters under the normal incidence.

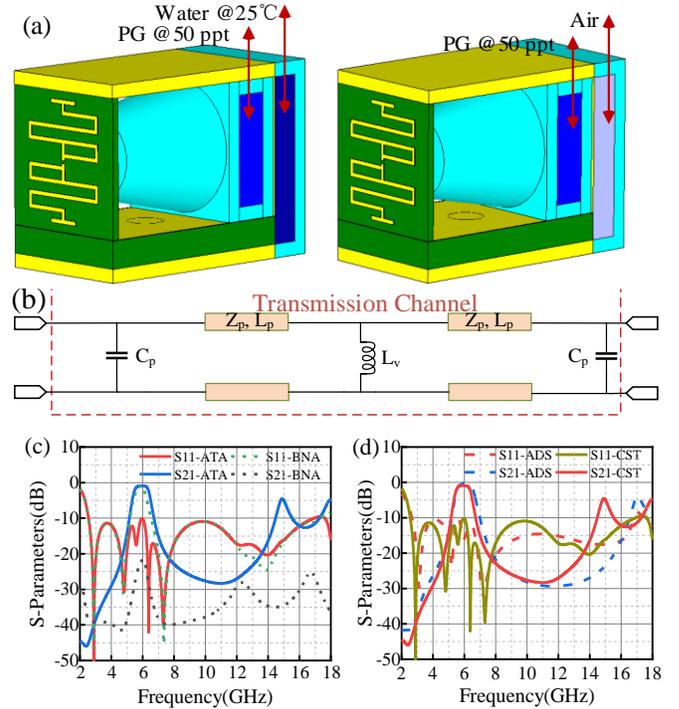

Fig. 8. Structure of the simplified FSR and its simulation results (a) 3-D view of the FSR under ARA mode (b) 3-D view of the FSR under ATA mode (c) ECM of the simplified FSR on BNA mode (d) Reflection and transmission coefficients of two modes (e) Comparison of S-Parameters by EM simulation (CST) and ECM under TE polarization for the FSR.

that the ECM can accurately describe the simulation results of water-based MA.

### C. Reconfigurable principle

Apart from using water as a resonator or dielectric loading in the design of a water antenna, water can also act as a reflector. Inspired by the water patch antenna, a water channel was put behind the FSR as a reflection layer. When the reflector is filled with water, the guided electromagnetic wave in the transmission channel will be reflected due to the significant impedance mismatch. To validate this method, a water layer with various heights was simulated in CST. The reflection coefficients as a function of height is shown in Fig. 7. As shown in Fig. 7, the reflection coefficient is greater than 0.8 at the yellow area in Fig. 7. When the water in the reflector is dried out, the slightly reflector's influence on the transmission path can be ignored. This demonstrates that the process of generating a reconfigurable layer using water is reliable.

## IV. EXPERIMENTAL MEASUREMENT AND RESULTS

Due to the difficulties of processing and manufacturing, a simplified model of transmission path is fabricated and measured to verify the correctness of the above design. As shows in Fig. 8(a) and (b), the reflection layer was filled with pure water under BNA mode and filed with air under the ATA mode. Compared to the original design, it replaces the SIR employed in the transmission path with a common media filled PPW form. The rest of the model is identical to the original design. In Fig. 8 (c), the ECM of a reduced structure is shown, which adjusts the corresponding section in relation to Fig. 6. The simplified FSR's simulated reflection and transmission coefficient results under transverse electric (TE)-polarized EM wave ($E$ along with $y$-direction) obtained by CST are shown in Fig. 8 (d). With a structure thickness of $0.159\ \lambda_c$, a fractional BW (FBW) of 147% is obtained in ATA mode from 2.55 GHz to 16.9 GHz. At the transmission band, the - 3 dB transmission BW is 18.5. In terms of BNA mode, S11 is less than -10 dB from 2.55 GHz to 5.9 GHz and 6.64 GHz to 17.3 GHz. At 5.91 GHz, a minimum insertion loss of 1 dB was obtained in the reflection band. The comparison of S-Parameters results between full-wave simulation and ECM is illustrated in Fig. 8 (e), where a good agreement can be observed. The difference above 12 GHz is mainly because the values of $C_p$ is assumed to be frequency independent in ECM, while these discontinuities vary in EM simulation at high frequencies.

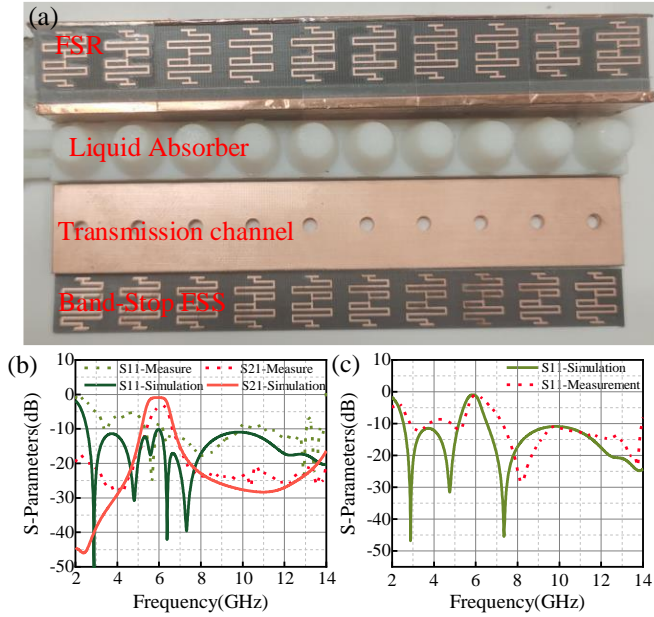

Fig. 9. (a) Fabricated prototype of the proposed FSR and (b) its simulated and measured S-parameters under the normal incidence under ATA mode. (c) BNA mode.

A prototype of this simplified FSR is fabricated and measured to verify the aforesaid design, as shown in Fig. 8. It contains of $10 \times 1$ unit cells along $x$- and $y$-directions, with a dimension of 100 mm × 14 mm and a thickness of 18.7 mm. The band-stop FSS is printed on a F4BM220 ($\varepsilon_r = 2.2\ and\ tan(\delta) = 0.001$) substrate with a thickness of 0.5 mm. The transmission path is constructed by metal painted holes on double-side copper clad laminate through PCB process and the substrate is F4BM220. 3-D printing technique was used to manufacture the container of MA and reflection layer. The transmission route and MA are joined using neutral silicone. The prototype is measured using a PPW setup, which has been validated and utilized in [28].

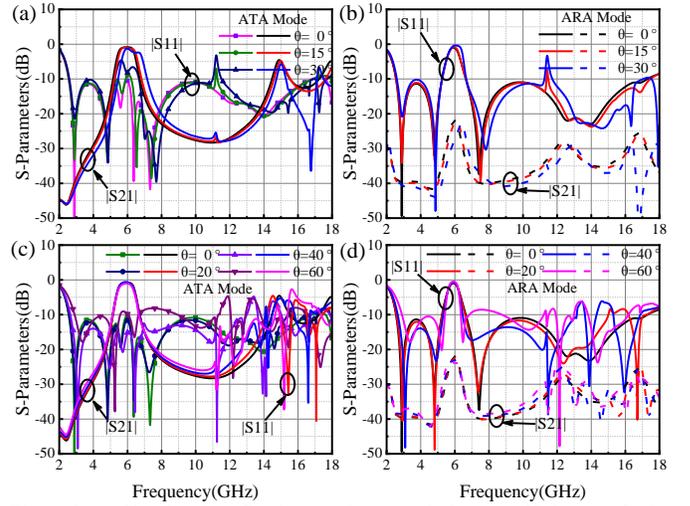

Fig. 10. Simulated reflection and transmission coefficients of the reconfigurable FSR under various incident angles. (a) ATA mode in the $xz$ plane ($\varphi = 0°$) (b) ARA mode at $xz$ plane ($\varphi = 0°$) (c) ATA mode at $yz$ plane ($\varphi = 90°$) and (d) ARA mode at $yz$ plane ($\varphi = 90°$).

The comparison between the simulated and measured frequency responses under the normal incidence is presented in Fig. 9. Due to the measurement setup in this paper can only handle up to 14 GHz, Fig. 8 only shows the test results from 2 to 14 GHz. Because of the ultra-wideband absorption qualities of the liquid absorber employed, the proposed FSR may still provide absorption band at higher frequencies. As shown in Fig. 9 (b) and (c), there was some mismatching between the simulation and measurement results of the proposed FSR in ATA mode. Mismatching might be produced by one or more of the following factors: 1) fabrication inaccuracy and mismatching in the assembling process, 2) uncertainty induced by foam absorbers used in testing devices and 3) frequency offset and extra loss caused by neutral silicone and glue used in assembly.

Because only the measurement under the normal incidence can be implemented in the PPW setup, the simulated S-parameters for oblique incidences are plotted in Fig. 10. It is noted that under the oblique incidence scanned in the $xz$ and $yz$ planes, the H-field is along the x-direction and the E-field is along the y-direction. It can be shown that our proposed FSR has a very stable frequency response, especially when scanned in the $yz$ plane at an oblique angle. A comparison with other reported FSR is made in Table I to validate the performance and advantages of our proposed 3-D liquid FSR structure. Clearly, the ultrawideband and switchable properties of our described FSR have been realized at the same time.

## V. CONCLUSION

In this paper, a switchable FSR with ultrawide absorption band was designed, fabricated and measured using liquid MWA and pure water as the reflection layer. The specific operating principle and mechanism have been demonstrated using ECM and multilayer medium theory. It is also worth noting that, to our understanding, the ECM of liquid MW was originally proposed in this paper. The ECM of a liquid microwave absorber was first proposed and employed in FSR design. It is worth noting that the ECM model is established by an algorithm rather than a complex parameter tuning procedure. As a result, this method has the advantages of fast convergence and high precision, and it may have practical value in absorber analysis. In ATA mode, our design example has a fractional absorption bandwidth of 146.8%, and in BNA mode, it has an absorption band of 2.5 to 4.6GHz and 5.7 to 16.5GHz. Good agreements among circuit analysis, simulation findings, and measurement results are finally obtained. Our proposed FSR is confined to a single polarization and suffers from the construction of the liquid absorber and reflection layer. Our design may be easier to assemble and apply if we use low-insertion 3D printing materials to complete the fabrication of the transmission path and switchable part. It can also be extended to dual-polarized designs with symmetrical or rotating structures. Our proposed design can be used in a shared-aperture antenna system to convert a broadband stealth radome into a reflector.

TABLE I
PERFORMANCE COMPARISON OF THE STATE-OF-THE-ART

| Ref. | Lower absorption FBW (%) | Higher absorption FBW (%) | 3D or 2D | Multifuction | Lossy principle |
|---|---|---|---|---|---|
| [7] | 92.8 | N.A. | 2D | N.A. | lumped resistance |
| [16] | 61.11 | 37.48 | 2D | N.A. | lumped resistance |
| [23] | 63.3 | 52.2 | 3D | N.A. | magnetic material |
| [27] | 130.3 | 58 | 3D | N.A. | magnetic material |
| [28] | 35.90 | 24.39 | 3D | N.A. | lumped resistance |
| | N.A. | 58.72 | 2D | Y | water |
| **This work** ATA mode | **49.17** | **94.54** | **3D** | **Y** | **PG aqueous solution** |
| BNA mode | **59.15** | **97.30** | | | |